\documentclass{article}
\pdfoutput=1

\PassOptionsToPackage{numbers, compress}{natbib}

\usepackage[final]{nips_2018}

\usepackage[utf8]{inputenc} %
\usepackage[T1]{fontenc}    %
\usepackage{hyperref}       %
\usepackage{url}            %
\usepackage{booktabs}       %
\usepackage{amsfonts}       %
\usepackage{nicefrac}       %
\usepackage{microtype}      %
\usepackage{graphicx}
\usepackage{pgf}
\title{Distilling Information from a Flood: A Possibility for the Use of Meta-Analysis and Systematic Review in Machine Learning Research}

\author{
  Peter Henderson\\
  Stanford University\\
  \texttt{phend@cs.stanford.edu}\\
  \And 
  Emma Brunksill\\
  Stanford University\\
  \texttt{ebrun@cs.stanford.edu}\\
}

\begin{document}

\maketitle

\begin{abstract}
The current flood of information in all areas of machine learning research, from computer vision to reinforcement learning, has made it difficult to make aggregate scientific inferences.
It can be challenging to distill a myriad of similar papers into a set of useful principles, to determine which new methodologies to use for a particular application, and to be confident that one has compared against all relevant related work when developing new ideas.
However, such a rapidly growing body of research literature is a problem that other fields have already faced -- in particular, medicine and epidemiology.
In those fields, systematic reviews and meta-analyses have been used exactly for dealing with these issues and it is not uncommon for entire journals to be dedicated to such analyses.
Here, we suggest the field of machine learning might similarly benefit from meta-analysis and systematic review, and we encourage further discussion and development along this direction.
\end{abstract}

\section{Introduction}

\begin{figure}[!htbp]
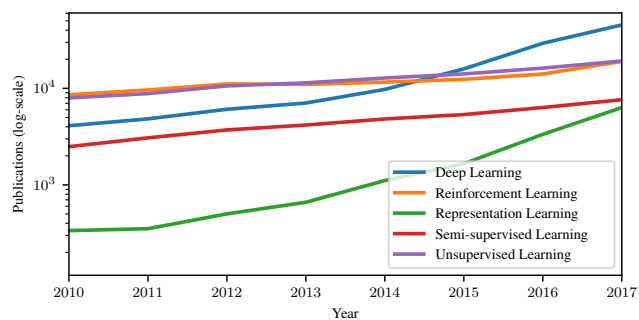

    \centering
\begingroup%
\makeatletter%
\resizebox{0.6\textwidth}{!}{
\begin{pgfpicture}%
\pgfpathrectangle{\pgfpointorigin}{\pgfqpoint{4.773765in}{2.386883in}}%
\pgfusepath{use as bounding box, clip}%
\begin{pgfscope}%
\pgfsetbuttcap%
\pgfsetmiterjoin%
\definecolor{currentfill}{rgb}{1.000000,1.000000,1.000000}%
\pgfsetfillcolor{currentfill}%
\pgfsetlinewidth{0.000000pt}%
\definecolor{currentstroke}{rgb}{1.000000,1.000000,1.000000}%
\pgfsetstrokecolor{currentstroke}%
\pgfsetdash{}{0pt}%
\pgfpathmoveto{\pgfqpoint{0.000000in}{0.000000in}}%
\pgfpathlineto{\pgfqpoint{4.773765in}{0.000000in}}%
\pgfpathlineto{\pgfqpoint{4.773765in}{2.386883in}}%
\pgfpathlineto{\pgfqpoint{0.000000in}{2.386883in}}%
\pgfpathclose%
\pgfusepath{fill}%
\end{pgfscope}%
\begin{pgfscope}%
\pgfsetbuttcap%
\pgfsetmiterjoin%
\definecolor{currentfill}{rgb}{1.000000,1.000000,1.000000}%
\pgfsetfillcolor{currentfill}%
\pgfsetlinewidth{0.000000pt}%
\definecolor{currentstroke}{rgb}{0.000000,0.000000,0.000000}%
\pgfsetstrokecolor{currentstroke}%
\pgfsetstrokeopacity{0.000000}%
\pgfsetdash{}{0pt}%
\pgfpathmoveto{\pgfqpoint{0.439815in}{0.350111in}}%
\pgfpathlineto{\pgfqpoint{4.655708in}{0.350111in}}%
\pgfpathlineto{\pgfqpoint{4.655708in}{2.351883in}}%
\pgfpathlineto{\pgfqpoint{0.439815in}{2.351883in}}%
\pgfpathclose%
\pgfusepath{fill}%
\end{pgfscope}%
\begin{pgfscope}%
\pgfsetbuttcap%
\pgfsetroundjoin%
\definecolor{currentfill}{rgb}{0.000000,0.000000,0.000000}%
\pgfsetfillcolor{currentfill}%
\pgfsetlinewidth{0.803000pt}%
\definecolor{currentstroke}{rgb}{0.000000,0.000000,0.000000}%
\pgfsetstrokecolor{currentstroke}%
\pgfsetdash{}{0pt}%
\pgfsys@defobject{currentmarker}{\pgfqpoint{0.000000in}{-0.048611in}}{\pgfqpoint{0.000000in}{0.000000in}}{%
\pgfpathmoveto{\pgfqpoint{0.000000in}{0.000000in}}%
\pgfpathlineto{\pgfqpoint{0.000000in}{-0.048611in}}%
\pgfusepath{stroke,fill}%
}%
\begin{pgfscope}%
\pgfsys@transformshift{0.439815in}{0.350111in}%
\pgfsys@useobject{currentmarker}{}%
\end{pgfscope}%
\end{pgfscope}%
\begin{pgfscope}%
\pgftext[x=0.439815in,y=0.252889in,,top]{\rmfamily\fontsize{8.000000}{9.600000}\selectfont \(\displaystyle 2010\)}%
\end{pgfscope}%
\begin{pgfscope}%
\pgfsetbuttcap%
\pgfsetroundjoin%
\definecolor{currentfill}{rgb}{0.000000,0.000000,0.000000}%
\pgfsetfillcolor{currentfill}%
\pgfsetlinewidth{0.803000pt}%
\definecolor{currentstroke}{rgb}{0.000000,0.000000,0.000000}%
\pgfsetstrokecolor{currentstroke}%
\pgfsetdash{}{0pt}%
\pgfsys@defobject{currentmarker}{\pgfqpoint{0.000000in}{-0.048611in}}{\pgfqpoint{0.000000in}{0.000000in}}{%
\pgfpathmoveto{\pgfqpoint{0.000000in}{0.000000in}}%
\pgfpathlineto{\pgfqpoint{0.000000in}{-0.048611in}}%
\pgfusepath{stroke,fill}%
}%
\begin{pgfscope}%
\pgfsys@transformshift{1.042086in}{0.350111in}%
\pgfsys@useobject{currentmarker}{}%
\end{pgfscope}%
\end{pgfscope}%
\begin{pgfscope}%
\pgftext[x=1.042086in,y=0.252889in,,top]{\rmfamily\fontsize{8.000000}{9.600000}\selectfont \(\displaystyle 2011\)}%
\end{pgfscope}%
\begin{pgfscope}%
\pgfsetbuttcap%
\pgfsetroundjoin%
\definecolor{currentfill}{rgb}{0.000000,0.000000,0.000000}%
\pgfsetfillcolor{currentfill}%
\pgfsetlinewidth{0.803000pt}%
\definecolor{currentstroke}{rgb}{0.000000,0.000000,0.000000}%
\pgfsetstrokecolor{currentstroke}%
\pgfsetdash{}{0pt}%
\pgfsys@defobject{currentmarker}{\pgfqpoint{0.000000in}{-0.048611in}}{\pgfqpoint{0.000000in}{0.000000in}}{%
\pgfpathmoveto{\pgfqpoint{0.000000in}{0.000000in}}%
\pgfpathlineto{\pgfqpoint{0.000000in}{-0.048611in}}%
\pgfusepath{stroke,fill}%
}%
\begin{pgfscope}%
\pgfsys@transformshift{1.644356in}{0.350111in}%
\pgfsys@useobject{currentmarker}{}%
\end{pgfscope}%
\end{pgfscope}%
\begin{pgfscope}%
\pgftext[x=1.644356in,y=0.252889in,,top]{\rmfamily\fontsize{8.000000}{9.600000}\selectfont \(\displaystyle 2012\)}%
\end{pgfscope}%
\begin{pgfscope}%
\pgfsetbuttcap%
\pgfsetroundjoin%
\definecolor{currentfill}{rgb}{0.000000,0.000000,0.000000}%
\pgfsetfillcolor{currentfill}%
\pgfsetlinewidth{0.803000pt}%
\definecolor{currentstroke}{rgb}{0.000000,0.000000,0.000000}%
\pgfsetstrokecolor{currentstroke}%
\pgfsetdash{}{0pt}%
\pgfsys@defobject{currentmarker}{\pgfqpoint{0.000000in}{-0.048611in}}{\pgfqpoint{0.000000in}{0.000000in}}{%
\pgfpathmoveto{\pgfqpoint{0.000000in}{0.000000in}}%
\pgfpathlineto{\pgfqpoint{0.000000in}{-0.048611in}}%
\pgfusepath{stroke,fill}%
}%
\begin{pgfscope}%
\pgfsys@transformshift{2.246626in}{0.350111in}%
\pgfsys@useobject{currentmarker}{}%
\end{pgfscope}%
\end{pgfscope}%
\begin{pgfscope}%
\pgftext[x=2.246626in,y=0.252889in,,top]{\rmfamily\fontsize{8.000000}{9.600000}\selectfont \(\displaystyle 2013\)}%
\end{pgfscope}%
\begin{pgfscope}%
\pgfsetbuttcap%
\pgfsetroundjoin%
\definecolor{currentfill}{rgb}{0.000000,0.000000,0.000000}%
\pgfsetfillcolor{currentfill}%
\pgfsetlinewidth{0.803000pt}%
\definecolor{currentstroke}{rgb}{0.000000,0.000000,0.000000}%
\pgfsetstrokecolor{currentstroke}%
\pgfsetdash{}{0pt}%
\pgfsys@defobject{currentmarker}{\pgfqpoint{0.000000in}{-0.048611in}}{\pgfqpoint{0.000000in}{0.000000in}}{%
\pgfpathmoveto{\pgfqpoint{0.000000in}{0.000000in}}%
\pgfpathlineto{\pgfqpoint{0.000000in}{-0.048611in}}%
\pgfusepath{stroke,fill}%
}%
\begin{pgfscope}%
\pgfsys@transformshift{2.848897in}{0.350111in}%
\pgfsys@useobject{currentmarker}{}%
\end{pgfscope}%
\end{pgfscope}%
\begin{pgfscope}%
\pgftext[x=2.848897in,y=0.252889in,,top]{\rmfamily\fontsize{8.000000}{9.600000}\selectfont \(\displaystyle 2014\)}%
\end{pgfscope}%
\begin{pgfscope}%
\pgfsetbuttcap%
\pgfsetroundjoin%
\definecolor{currentfill}{rgb}{0.000000,0.000000,0.000000}%
\pgfsetfillcolor{currentfill}%
\pgfsetlinewidth{0.803000pt}%
\definecolor{currentstroke}{rgb}{0.000000,0.000000,0.000000}%
\pgfsetstrokecolor{currentstroke}%
\pgfsetdash{}{0pt}%
\pgfsys@defobject{currentmarker}{\pgfqpoint{0.000000in}{-0.048611in}}{\pgfqpoint{0.000000in}{0.000000in}}{%
\pgfpathmoveto{\pgfqpoint{0.000000in}{0.000000in}}%
\pgfpathlineto{\pgfqpoint{0.000000in}{-0.048611in}}%
\pgfusepath{stroke,fill}%
}%
\begin{pgfscope}%
\pgfsys@transformshift{3.451167in}{0.350111in}%
\pgfsys@useobject{currentmarker}{}%
\end{pgfscope}%
\end{pgfscope}%
\begin{pgfscope}%
\pgftext[x=3.451167in,y=0.252889in,,top]{\rmfamily\fontsize{8.000000}{9.600000}\selectfont \(\displaystyle 2015\)}%
\end{pgfscope}%
\begin{pgfscope}%
\pgfsetbuttcap%
\pgfsetroundjoin%
\definecolor{currentfill}{rgb}{0.000000,0.000000,0.000000}%
\pgfsetfillcolor{currentfill}%
\pgfsetlinewidth{0.803000pt}%
\definecolor{currentstroke}{rgb}{0.000000,0.000000,0.000000}%
\pgfsetstrokecolor{currentstroke}%
\pgfsetdash{}{0pt}%
\pgfsys@defobject{currentmarker}{\pgfqpoint{0.000000in}{-0.048611in}}{\pgfqpoint{0.000000in}{0.000000in}}{%
\pgfpathmoveto{\pgfqpoint{0.000000in}{0.000000in}}%
\pgfpathlineto{\pgfqpoint{0.000000in}{-0.048611in}}%
\pgfusepath{stroke,fill}%
}%
\begin{pgfscope}%
\pgfsys@transformshift{4.053437in}{0.350111in}%
\pgfsys@useobject{currentmarker}{}%
\end{pgfscope}%
\end{pgfscope}%
\begin{pgfscope}%
\pgftext[x=4.053437in,y=0.252889in,,top]{\rmfamily\fontsize{8.000000}{9.600000}\selectfont \(\displaystyle 2016\)}%
\end{pgfscope}%
\begin{pgfscope}%
\pgfsetbuttcap%
\pgfsetroundjoin%
\definecolor{currentfill}{rgb}{0.000000,0.000000,0.000000}%
\pgfsetfillcolor{currentfill}%
\pgfsetlinewidth{0.803000pt}%
\definecolor{currentstroke}{rgb}{0.000000,0.000000,0.000000}%
\pgfsetstrokecolor{currentstroke}%
\pgfsetdash{}{0pt}%
\pgfsys@defobject{currentmarker}{\pgfqpoint{0.000000in}{-0.048611in}}{\pgfqpoint{0.000000in}{0.000000in}}{%
\pgfpathmoveto{\pgfqpoint{0.000000in}{0.000000in}}%
\pgfpathlineto{\pgfqpoint{0.000000in}{-0.048611in}}%
\pgfusepath{stroke,fill}%
}%
\begin{pgfscope}%
\pgfsys@transformshift{4.655708in}{0.350111in}%
\pgfsys@useobject{currentmarker}{}%
\end{pgfscope}%
\end{pgfscope}%
\begin{pgfscope}%
\pgftext[x=4.655708in,y=0.252889in,,top]{\rmfamily\fontsize{8.000000}{9.600000}\selectfont \(\displaystyle 2017\)}%
\end{pgfscope}%
\begin{pgfscope}%
\pgftext[x=2.547762in,y=0.098667in,,top]{\rmfamily\fontsize{8.000000}{9.600000}\selectfont Year}%
\end{pgfscope}%
\begin{pgfscope}%
\pgfsetbuttcap%
\pgfsetroundjoin%
\definecolor{currentfill}{rgb}{0.000000,0.000000,0.000000}%
\pgfsetfillcolor{currentfill}%
\pgfsetlinewidth{0.803000pt}%
\definecolor{currentstroke}{rgb}{0.000000,0.000000,0.000000}%
\pgfsetstrokecolor{currentstroke}%
\pgfsetdash{}{0pt}%
\pgfsys@defobject{currentmarker}{\pgfqpoint{-0.048611in}{0.000000in}}{\pgfqpoint{0.000000in}{0.000000in}}{%
\pgfpathmoveto{\pgfqpoint{0.000000in}{0.000000in}}%
\pgfpathlineto{\pgfqpoint{-0.048611in}{0.000000in}}%
\pgfusepath{stroke,fill}%
}%
\begin{pgfscope}%
\pgfsys@transformshift{0.439815in}{1.039814in}%
\pgfsys@useobject{currentmarker}{}%
\end{pgfscope}%
\end{pgfscope}%
\begin{pgfscope}%
\pgftext[x=0.166666in,y=1.000661in,left,base]{\rmfamily\fontsize{8.000000}{9.600000}\selectfont \(\displaystyle 10^{3}\)}%
\end{pgfscope}%
\begin{pgfscope}%
\pgfsetbuttcap%
\pgfsetroundjoin%
\definecolor{currentfill}{rgb}{0.000000,0.000000,0.000000}%
\pgfsetfillcolor{currentfill}%
\pgfsetlinewidth{0.803000pt}%
\definecolor{currentstroke}{rgb}{0.000000,0.000000,0.000000}%
\pgfsetstrokecolor{currentstroke}%
\pgfsetdash{}{0pt}%
\pgfsys@defobject{currentmarker}{\pgfqpoint{-0.048611in}{0.000000in}}{\pgfqpoint{0.000000in}{0.000000in}}{%
\pgfpathmoveto{\pgfqpoint{0.000000in}{0.000000in}}%
\pgfpathlineto{\pgfqpoint{-0.048611in}{0.000000in}}%
\pgfusepath{stroke,fill}%
}%
\begin{pgfscope}%
\pgfsys@transformshift{0.439815in}{1.776711in}%
\pgfsys@useobject{currentmarker}{}%
\end{pgfscope}%
\end{pgfscope}%
\begin{pgfscope}%
\pgftext[x=0.166666in,y=1.737558in,left,base]{\rmfamily\fontsize{8.000000}{9.600000}\selectfont \(\displaystyle 10^{4}\)}%
\end{pgfscope}%
\begin{pgfscope}%
\pgfsetbuttcap%
\pgfsetroundjoin%
\definecolor{currentfill}{rgb}{0.000000,0.000000,0.000000}%
\pgfsetfillcolor{currentfill}%
\pgfsetlinewidth{0.602250pt}%
\definecolor{currentstroke}{rgb}{0.000000,0.000000,0.000000}%
\pgfsetstrokecolor{currentstroke}%
\pgfsetdash{}{0pt}%
\pgfsys@defobject{currentmarker}{\pgfqpoint{-0.027778in}{0.000000in}}{\pgfqpoint{0.000000in}{0.000000in}}{%
\pgfpathmoveto{\pgfqpoint{0.000000in}{0.000000in}}%
\pgfpathlineto{\pgfqpoint{-0.027778in}{0.000000in}}%
\pgfusepath{stroke,fill}%
}%
\begin{pgfscope}%
\pgfsys@transformshift{0.439815in}{0.524745in}%
\pgfsys@useobject{currentmarker}{}%
\end{pgfscope}%
\end{pgfscope}%
\begin{pgfscope}%
\pgfsetbuttcap%
\pgfsetroundjoin%
\definecolor{currentfill}{rgb}{0.000000,0.000000,0.000000}%
\pgfsetfillcolor{currentfill}%
\pgfsetlinewidth{0.602250pt}%
\definecolor{currentstroke}{rgb}{0.000000,0.000000,0.000000}%
\pgfsetstrokecolor{currentstroke}%
\pgfsetdash{}{0pt}%
\pgfsys@defobject{currentmarker}{\pgfqpoint{-0.027778in}{0.000000in}}{\pgfqpoint{0.000000in}{0.000000in}}{%
\pgfpathmoveto{\pgfqpoint{0.000000in}{0.000000in}}%
\pgfpathlineto{\pgfqpoint{-0.027778in}{0.000000in}}%
\pgfusepath{stroke,fill}%
}%
\begin{pgfscope}%
\pgfsys@transformshift{0.439815in}{0.654506in}%
\pgfsys@useobject{currentmarker}{}%
\end{pgfscope}%
\end{pgfscope}%
\begin{pgfscope}%
\pgfsetbuttcap%
\pgfsetroundjoin%
\definecolor{currentfill}{rgb}{0.000000,0.000000,0.000000}%
\pgfsetfillcolor{currentfill}%
\pgfsetlinewidth{0.602250pt}%
\definecolor{currentstroke}{rgb}{0.000000,0.000000,0.000000}%
\pgfsetstrokecolor{currentstroke}%
\pgfsetdash{}{0pt}%
\pgfsys@defobject{currentmarker}{\pgfqpoint{-0.027778in}{0.000000in}}{\pgfqpoint{0.000000in}{0.000000in}}{%
\pgfpathmoveto{\pgfqpoint{0.000000in}{0.000000in}}%
\pgfpathlineto{\pgfqpoint{-0.027778in}{0.000000in}}%
\pgfusepath{stroke,fill}%
}%
\begin{pgfscope}%
\pgfsys@transformshift{0.439815in}{0.746573in}%
\pgfsys@useobject{currentmarker}{}%
\end{pgfscope}%
\end{pgfscope}%
\begin{pgfscope}%
\pgfsetbuttcap%
\pgfsetroundjoin%
\definecolor{currentfill}{rgb}{0.000000,0.000000,0.000000}%
\pgfsetfillcolor{currentfill}%
\pgfsetlinewidth{0.602250pt}%
\definecolor{currentstroke}{rgb}{0.000000,0.000000,0.000000}%
\pgfsetstrokecolor{currentstroke}%
\pgfsetdash{}{0pt}%
\pgfsys@defobject{currentmarker}{\pgfqpoint{-0.027778in}{0.000000in}}{\pgfqpoint{0.000000in}{0.000000in}}{%
\pgfpathmoveto{\pgfqpoint{0.000000in}{0.000000in}}%
\pgfpathlineto{\pgfqpoint{-0.027778in}{0.000000in}}%
\pgfusepath{stroke,fill}%
}%
\begin{pgfscope}%
\pgfsys@transformshift{0.439815in}{0.817986in}%
\pgfsys@useobject{currentmarker}{}%
\end{pgfscope}%
\end{pgfscope}%
\begin{pgfscope}%
\pgfsetbuttcap%
\pgfsetroundjoin%
\definecolor{currentfill}{rgb}{0.000000,0.000000,0.000000}%
\pgfsetfillcolor{currentfill}%
\pgfsetlinewidth{0.602250pt}%
\definecolor{currentstroke}{rgb}{0.000000,0.000000,0.000000}%
\pgfsetstrokecolor{currentstroke}%
\pgfsetdash{}{0pt}%
\pgfsys@defobject{currentmarker}{\pgfqpoint{-0.027778in}{0.000000in}}{\pgfqpoint{0.000000in}{0.000000in}}{%
\pgfpathmoveto{\pgfqpoint{0.000000in}{0.000000in}}%
\pgfpathlineto{\pgfqpoint{-0.027778in}{0.000000in}}%
\pgfusepath{stroke,fill}%
}%
\begin{pgfscope}%
\pgfsys@transformshift{0.439815in}{0.876334in}%
\pgfsys@useobject{currentmarker}{}%
\end{pgfscope}%
\end{pgfscope}%
\begin{pgfscope}%
\pgfsetbuttcap%
\pgfsetroundjoin%
\definecolor{currentfill}{rgb}{0.000000,0.000000,0.000000}%
\pgfsetfillcolor{currentfill}%
\pgfsetlinewidth{0.602250pt}%
\definecolor{currentstroke}{rgb}{0.000000,0.000000,0.000000}%
\pgfsetstrokecolor{currentstroke}%
\pgfsetdash{}{0pt}%
\pgfsys@defobject{currentmarker}{\pgfqpoint{-0.027778in}{0.000000in}}{\pgfqpoint{0.000000in}{0.000000in}}{%
\pgfpathmoveto{\pgfqpoint{0.000000in}{0.000000in}}%
\pgfpathlineto{\pgfqpoint{-0.027778in}{0.000000in}}%
\pgfusepath{stroke,fill}%
}%
\begin{pgfscope}%
\pgfsys@transformshift{0.439815in}{0.925667in}%
\pgfsys@useobject{currentmarker}{}%
\end{pgfscope}%
\end{pgfscope}%
\begin{pgfscope}%
\pgfsetbuttcap%
\pgfsetroundjoin%
\definecolor{currentfill}{rgb}{0.000000,0.000000,0.000000}%
\pgfsetfillcolor{currentfill}%
\pgfsetlinewidth{0.602250pt}%
\definecolor{currentstroke}{rgb}{0.000000,0.000000,0.000000}%
\pgfsetstrokecolor{currentstroke}%
\pgfsetdash{}{0pt}%
\pgfsys@defobject{currentmarker}{\pgfqpoint{-0.027778in}{0.000000in}}{\pgfqpoint{0.000000in}{0.000000in}}{%
\pgfpathmoveto{\pgfqpoint{0.000000in}{0.000000in}}%
\pgfpathlineto{\pgfqpoint{-0.027778in}{0.000000in}}%
\pgfusepath{stroke,fill}%
}%
\begin{pgfscope}%
\pgfsys@transformshift{0.439815in}{0.968401in}%
\pgfsys@useobject{currentmarker}{}%
\end{pgfscope}%
\end{pgfscope}%
\begin{pgfscope}%
\pgfsetbuttcap%
\pgfsetroundjoin%
\definecolor{currentfill}{rgb}{0.000000,0.000000,0.000000}%
\pgfsetfillcolor{currentfill}%
\pgfsetlinewidth{0.602250pt}%
\definecolor{currentstroke}{rgb}{0.000000,0.000000,0.000000}%
\pgfsetstrokecolor{currentstroke}%
\pgfsetdash{}{0pt}%
\pgfsys@defobject{currentmarker}{\pgfqpoint{-0.027778in}{0.000000in}}{\pgfqpoint{0.000000in}{0.000000in}}{%
\pgfpathmoveto{\pgfqpoint{0.000000in}{0.000000in}}%
\pgfpathlineto{\pgfqpoint{-0.027778in}{0.000000in}}%
\pgfusepath{stroke,fill}%
}%
\begin{pgfscope}%
\pgfsys@transformshift{0.439815in}{1.006095in}%
\pgfsys@useobject{currentmarker}{}%
\end{pgfscope}%
\end{pgfscope}%
\begin{pgfscope}%
\pgfsetbuttcap%
\pgfsetroundjoin%
\definecolor{currentfill}{rgb}{0.000000,0.000000,0.000000}%
\pgfsetfillcolor{currentfill}%
\pgfsetlinewidth{0.602250pt}%
\definecolor{currentstroke}{rgb}{0.000000,0.000000,0.000000}%
\pgfsetstrokecolor{currentstroke}%
\pgfsetdash{}{0pt}%
\pgfsys@defobject{currentmarker}{\pgfqpoint{-0.027778in}{0.000000in}}{\pgfqpoint{0.000000in}{0.000000in}}{%
\pgfpathmoveto{\pgfqpoint{0.000000in}{0.000000in}}%
\pgfpathlineto{\pgfqpoint{-0.027778in}{0.000000in}}%
\pgfusepath{stroke,fill}%
}%
\begin{pgfscope}%
\pgfsys@transformshift{0.439815in}{1.261642in}%
\pgfsys@useobject{currentmarker}{}%
\end{pgfscope}%
\end{pgfscope}%
\begin{pgfscope}%
\pgfsetbuttcap%
\pgfsetroundjoin%
\definecolor{currentfill}{rgb}{0.000000,0.000000,0.000000}%
\pgfsetfillcolor{currentfill}%
\pgfsetlinewidth{0.602250pt}%
\definecolor{currentstroke}{rgb}{0.000000,0.000000,0.000000}%
\pgfsetstrokecolor{currentstroke}%
\pgfsetdash{}{0pt}%
\pgfsys@defobject{currentmarker}{\pgfqpoint{-0.027778in}{0.000000in}}{\pgfqpoint{0.000000in}{0.000000in}}{%
\pgfpathmoveto{\pgfqpoint{0.000000in}{0.000000in}}%
\pgfpathlineto{\pgfqpoint{-0.027778in}{0.000000in}}%
\pgfusepath{stroke,fill}%
}%
\begin{pgfscope}%
\pgfsys@transformshift{0.439815in}{1.391403in}%
\pgfsys@useobject{currentmarker}{}%
\end{pgfscope}%
\end{pgfscope}%
\begin{pgfscope}%
\pgfsetbuttcap%
\pgfsetroundjoin%
\definecolor{currentfill}{rgb}{0.000000,0.000000,0.000000}%
\pgfsetfillcolor{currentfill}%
\pgfsetlinewidth{0.602250pt}%
\definecolor{currentstroke}{rgb}{0.000000,0.000000,0.000000}%
\pgfsetstrokecolor{currentstroke}%
\pgfsetdash{}{0pt}%
\pgfsys@defobject{currentmarker}{\pgfqpoint{-0.027778in}{0.000000in}}{\pgfqpoint{0.000000in}{0.000000in}}{%
\pgfpathmoveto{\pgfqpoint{0.000000in}{0.000000in}}%
\pgfpathlineto{\pgfqpoint{-0.027778in}{0.000000in}}%
\pgfusepath{stroke,fill}%
}%
\begin{pgfscope}%
\pgfsys@transformshift{0.439815in}{1.483470in}%
\pgfsys@useobject{currentmarker}{}%
\end{pgfscope}%
\end{pgfscope}%
\begin{pgfscope}%
\pgfsetbuttcap%
\pgfsetroundjoin%
\definecolor{currentfill}{rgb}{0.000000,0.000000,0.000000}%
\pgfsetfillcolor{currentfill}%
\pgfsetlinewidth{0.602250pt}%
\definecolor{currentstroke}{rgb}{0.000000,0.000000,0.000000}%
\pgfsetstrokecolor{currentstroke}%
\pgfsetdash{}{0pt}%
\pgfsys@defobject{currentmarker}{\pgfqpoint{-0.027778in}{0.000000in}}{\pgfqpoint{0.000000in}{0.000000in}}{%
\pgfpathmoveto{\pgfqpoint{0.000000in}{0.000000in}}%
\pgfpathlineto{\pgfqpoint{-0.027778in}{0.000000in}}%
\pgfusepath{stroke,fill}%
}%
\begin{pgfscope}%
\pgfsys@transformshift{0.439815in}{1.554883in}%
\pgfsys@useobject{currentmarker}{}%
\end{pgfscope}%
\end{pgfscope}%
\begin{pgfscope}%
\pgfsetbuttcap%
\pgfsetroundjoin%
\definecolor{currentfill}{rgb}{0.000000,0.000000,0.000000}%
\pgfsetfillcolor{currentfill}%
\pgfsetlinewidth{0.602250pt}%
\definecolor{currentstroke}{rgb}{0.000000,0.000000,0.000000}%
\pgfsetstrokecolor{currentstroke}%
\pgfsetdash{}{0pt}%
\pgfsys@defobject{currentmarker}{\pgfqpoint{-0.027778in}{0.000000in}}{\pgfqpoint{0.000000in}{0.000000in}}{%
\pgfpathmoveto{\pgfqpoint{0.000000in}{0.000000in}}%
\pgfpathlineto{\pgfqpoint{-0.027778in}{0.000000in}}%
\pgfusepath{stroke,fill}%
}%
\begin{pgfscope}%
\pgfsys@transformshift{0.439815in}{1.613231in}%
\pgfsys@useobject{currentmarker}{}%
\end{pgfscope}%
\end{pgfscope}%
\begin{pgfscope}%
\pgfsetbuttcap%
\pgfsetroundjoin%
\definecolor{currentfill}{rgb}{0.000000,0.000000,0.000000}%
\pgfsetfillcolor{currentfill}%
\pgfsetlinewidth{0.602250pt}%
\definecolor{currentstroke}{rgb}{0.000000,0.000000,0.000000}%
\pgfsetstrokecolor{currentstroke}%
\pgfsetdash{}{0pt}%
\pgfsys@defobject{currentmarker}{\pgfqpoint{-0.027778in}{0.000000in}}{\pgfqpoint{0.000000in}{0.000000in}}{%
\pgfpathmoveto{\pgfqpoint{0.000000in}{0.000000in}}%
\pgfpathlineto{\pgfqpoint{-0.027778in}{0.000000in}}%
\pgfusepath{stroke,fill}%
}%
\begin{pgfscope}%
\pgfsys@transformshift{0.439815in}{1.662564in}%
\pgfsys@useobject{currentmarker}{}%
\end{pgfscope}%
\end{pgfscope}%
\begin{pgfscope}%
\pgfsetbuttcap%
\pgfsetroundjoin%
\definecolor{currentfill}{rgb}{0.000000,0.000000,0.000000}%
\pgfsetfillcolor{currentfill}%
\pgfsetlinewidth{0.602250pt}%
\definecolor{currentstroke}{rgb}{0.000000,0.000000,0.000000}%
\pgfsetstrokecolor{currentstroke}%
\pgfsetdash{}{0pt}%
\pgfsys@defobject{currentmarker}{\pgfqpoint{-0.027778in}{0.000000in}}{\pgfqpoint{0.000000in}{0.000000in}}{%
\pgfpathmoveto{\pgfqpoint{0.000000in}{0.000000in}}%
\pgfpathlineto{\pgfqpoint{-0.027778in}{0.000000in}}%
\pgfusepath{stroke,fill}%
}%
\begin{pgfscope}%
\pgfsys@transformshift{0.439815in}{1.705298in}%
\pgfsys@useobject{currentmarker}{}%
\end{pgfscope}%
\end{pgfscope}%
\begin{pgfscope}%
\pgfsetbuttcap%
\pgfsetroundjoin%
\definecolor{currentfill}{rgb}{0.000000,0.000000,0.000000}%
\pgfsetfillcolor{currentfill}%
\pgfsetlinewidth{0.602250pt}%
\definecolor{currentstroke}{rgb}{0.000000,0.000000,0.000000}%
\pgfsetstrokecolor{currentstroke}%
\pgfsetdash{}{0pt}%
\pgfsys@defobject{currentmarker}{\pgfqpoint{-0.027778in}{0.000000in}}{\pgfqpoint{0.000000in}{0.000000in}}{%
\pgfpathmoveto{\pgfqpoint{0.000000in}{0.000000in}}%
\pgfpathlineto{\pgfqpoint{-0.027778in}{0.000000in}}%
\pgfusepath{stroke,fill}%
}%
\begin{pgfscope}%
\pgfsys@transformshift{0.439815in}{1.742992in}%
\pgfsys@useobject{currentmarker}{}%
\end{pgfscope}%
\end{pgfscope}%
\begin{pgfscope}%
\pgfsetbuttcap%
\pgfsetroundjoin%
\definecolor{currentfill}{rgb}{0.000000,0.000000,0.000000}%
\pgfsetfillcolor{currentfill}%
\pgfsetlinewidth{0.602250pt}%
\definecolor{currentstroke}{rgb}{0.000000,0.000000,0.000000}%
\pgfsetstrokecolor{currentstroke}%
\pgfsetdash{}{0pt}%
\pgfsys@defobject{currentmarker}{\pgfqpoint{-0.027778in}{0.000000in}}{\pgfqpoint{0.000000in}{0.000000in}}{%
\pgfpathmoveto{\pgfqpoint{0.000000in}{0.000000in}}%
\pgfpathlineto{\pgfqpoint{-0.027778in}{0.000000in}}%
\pgfusepath{stroke,fill}%
}%
\begin{pgfscope}%
\pgfsys@transformshift{0.439815in}{1.998539in}%
\pgfsys@useobject{currentmarker}{}%
\end{pgfscope}%
\end{pgfscope}%
\begin{pgfscope}%
\pgfsetbuttcap%
\pgfsetroundjoin%
\definecolor{currentfill}{rgb}{0.000000,0.000000,0.000000}%
\pgfsetfillcolor{currentfill}%
\pgfsetlinewidth{0.602250pt}%
\definecolor{currentstroke}{rgb}{0.000000,0.000000,0.000000}%
\pgfsetstrokecolor{currentstroke}%
\pgfsetdash{}{0pt}%
\pgfsys@defobject{currentmarker}{\pgfqpoint{-0.027778in}{0.000000in}}{\pgfqpoint{0.000000in}{0.000000in}}{%
\pgfpathmoveto{\pgfqpoint{0.000000in}{0.000000in}}%
\pgfpathlineto{\pgfqpoint{-0.027778in}{0.000000in}}%
\pgfusepath{stroke,fill}%
}%
\begin{pgfscope}%
\pgfsys@transformshift{0.439815in}{2.128300in}%
\pgfsys@useobject{currentmarker}{}%
\end{pgfscope}%
\end{pgfscope}%
\begin{pgfscope}%
\pgfsetbuttcap%
\pgfsetroundjoin%
\definecolor{currentfill}{rgb}{0.000000,0.000000,0.000000}%
\pgfsetfillcolor{currentfill}%
\pgfsetlinewidth{0.602250pt}%
\definecolor{currentstroke}{rgb}{0.000000,0.000000,0.000000}%
\pgfsetstrokecolor{currentstroke}%
\pgfsetdash{}{0pt}%
\pgfsys@defobject{currentmarker}{\pgfqpoint{-0.027778in}{0.000000in}}{\pgfqpoint{0.000000in}{0.000000in}}{%
\pgfpathmoveto{\pgfqpoint{0.000000in}{0.000000in}}%
\pgfpathlineto{\pgfqpoint{-0.027778in}{0.000000in}}%
\pgfusepath{stroke,fill}%
}%
\begin{pgfscope}%
\pgfsys@transformshift{0.439815in}{2.220367in}%
\pgfsys@useobject{currentmarker}{}%
\end{pgfscope}%
\end{pgfscope}%
\begin{pgfscope}%
\pgfsetbuttcap%
\pgfsetroundjoin%
\definecolor{currentfill}{rgb}{0.000000,0.000000,0.000000}%
\pgfsetfillcolor{currentfill}%
\pgfsetlinewidth{0.602250pt}%
\definecolor{currentstroke}{rgb}{0.000000,0.000000,0.000000}%
\pgfsetstrokecolor{currentstroke}%
\pgfsetdash{}{0pt}%
\pgfsys@defobject{currentmarker}{\pgfqpoint{-0.027778in}{0.000000in}}{\pgfqpoint{0.000000in}{0.000000in}}{%
\pgfpathmoveto{\pgfqpoint{0.000000in}{0.000000in}}%
\pgfpathlineto{\pgfqpoint{-0.027778in}{0.000000in}}%
\pgfusepath{stroke,fill}%
}%
\begin{pgfscope}%
\pgfsys@transformshift{0.439815in}{2.291779in}%
\pgfsys@useobject{currentmarker}{}%
\end{pgfscope}%
\end{pgfscope}%
\begin{pgfscope}%
\pgfsetbuttcap%
\pgfsetroundjoin%
\definecolor{currentfill}{rgb}{0.000000,0.000000,0.000000}%
\pgfsetfillcolor{currentfill}%
\pgfsetlinewidth{0.602250pt}%
\definecolor{currentstroke}{rgb}{0.000000,0.000000,0.000000}%
\pgfsetstrokecolor{currentstroke}%
\pgfsetdash{}{0pt}%
\pgfsys@defobject{currentmarker}{\pgfqpoint{-0.027778in}{0.000000in}}{\pgfqpoint{0.000000in}{0.000000in}}{%
\pgfpathmoveto{\pgfqpoint{0.000000in}{0.000000in}}%
\pgfpathlineto{\pgfqpoint{-0.027778in}{0.000000in}}%
\pgfusepath{stroke,fill}%
}%
\begin{pgfscope}%
\pgfsys@transformshift{0.439815in}{2.350128in}%
\pgfsys@useobject{currentmarker}{}%
\end{pgfscope}%
\end{pgfscope}%
\begin{pgfscope}%
\pgftext[x=0.111111in,y=1.350997in,,bottom,rotate=90.000000]{\rmfamily\fontsize{8.000000}{9.600000}\selectfont Publications (log-scale)}%
\end{pgfscope}%
\begin{pgfscope}%
\pgfpathrectangle{\pgfqpoint{0.439815in}{0.350111in}}{\pgfqpoint{4.215892in}{2.001772in}} %
\pgfusepath{clip}%
\pgfsetrectcap%
\pgfsetroundjoin%
\pgfsetlinewidth{2.007500pt}%
\definecolor{currentstroke}{rgb}{0.121569,0.466667,0.705882}%
\pgfsetstrokecolor{currentstroke}%
\pgfsetdash{}{0pt}%
\pgfpathmoveto{\pgfqpoint{0.429815in}{1.490622in}}%
\pgfpathlineto{\pgfqpoint{0.439815in}{1.491372in}}%
\pgfpathlineto{\pgfqpoint{1.042086in}{1.543812in}}%
\pgfpathlineto{\pgfqpoint{1.644356in}{1.616943in}}%
\pgfpathlineto{\pgfqpoint{2.246626in}{1.664842in}}%
\pgfpathlineto{\pgfqpoint{2.848897in}{1.768608in}}%
\pgfpathlineto{\pgfqpoint{3.451167in}{1.925119in}}%
\pgfpathlineto{\pgfqpoint{4.053437in}{2.120744in}}%
\pgfpathlineto{\pgfqpoint{4.655708in}{2.260893in}}%
\pgfusepath{stroke}%
\end{pgfscope}%
\begin{pgfscope}%
\pgfpathrectangle{\pgfqpoint{0.439815in}{0.350111in}}{\pgfqpoint{4.215892in}{2.001772in}} %
\pgfusepath{clip}%
\pgfsetrectcap%
\pgfsetroundjoin%
\pgfsetlinewidth{2.007500pt}%
\definecolor{currentstroke}{rgb}{1.000000,0.498039,0.054902}%
\pgfsetstrokecolor{currentstroke}%
\pgfsetdash{}{0pt}%
\pgfpathmoveto{\pgfqpoint{0.429815in}{1.727365in}}%
\pgfpathlineto{\pgfqpoint{0.439815in}{1.727698in}}%
\pgfpathlineto{\pgfqpoint{1.042086in}{1.764312in}}%
\pgfpathlineto{\pgfqpoint{1.644356in}{1.810109in}}%
\pgfpathlineto{\pgfqpoint{2.246626in}{1.807213in}}%
\pgfpathlineto{\pgfqpoint{2.848897in}{1.824209in}}%
\pgfpathlineto{\pgfqpoint{3.451167in}{1.845553in}}%
\pgfpathlineto{\pgfqpoint{4.053437in}{1.886670in}}%
\pgfpathlineto{\pgfqpoint{4.655708in}{1.982123in}}%
\pgfusepath{stroke}%
\end{pgfscope}%
\begin{pgfscope}%
\pgfpathrectangle{\pgfqpoint{0.439815in}{0.350111in}}{\pgfqpoint{4.215892in}{2.001772in}} %
\pgfusepath{clip}%
\pgfsetrectcap%
\pgfsetroundjoin%
\pgfsetlinewidth{2.007500pt}%
\definecolor{currentstroke}{rgb}{0.172549,0.627451,0.172549}%
\pgfsetstrokecolor{currentstroke}%
\pgfsetdash{}{0pt}%
\pgfpathmoveto{\pgfqpoint{0.429815in}{0.690065in}}%
\pgfpathlineto{\pgfqpoint{0.439815in}{0.690775in}}%
\pgfpathlineto{\pgfqpoint{1.042086in}{0.705663in}}%
\pgfpathlineto{\pgfqpoint{1.644356in}{0.818625in}}%
\pgfpathlineto{\pgfqpoint{2.246626in}{0.907321in}}%
\pgfpathlineto{\pgfqpoint{2.848897in}{1.073212in}}%
\pgfpathlineto{\pgfqpoint{3.451167in}{1.202011in}}%
\pgfpathlineto{\pgfqpoint{4.053437in}{1.425761in}}%
\pgfpathlineto{\pgfqpoint{4.655708in}{1.629860in}}%
\pgfusepath{stroke}%
\end{pgfscope}%
\begin{pgfscope}%
\pgfpathrectangle{\pgfqpoint{0.439815in}{0.350111in}}{\pgfqpoint{4.215892in}{2.001772in}} %
\pgfusepath{clip}%
\pgfsetrectcap%
\pgfsetroundjoin%
\pgfsetlinewidth{2.007500pt}%
\definecolor{currentstroke}{rgb}{0.839216,0.152941,0.156863}%
\pgfsetstrokecolor{currentstroke}%
\pgfsetdash{}{0pt}%
\pgfpathmoveto{\pgfqpoint{0.429815in}{1.330867in}}%
\pgfpathlineto{\pgfqpoint{0.439815in}{1.331772in}}%
\pgfpathlineto{\pgfqpoint{1.042086in}{1.398784in}}%
\pgfpathlineto{\pgfqpoint{1.644356in}{1.459384in}}%
\pgfpathlineto{\pgfqpoint{2.246626in}{1.496790in}}%
\pgfpathlineto{\pgfqpoint{2.848897in}{1.543149in}}%
\pgfpathlineto{\pgfqpoint{3.451167in}{1.576535in}}%
\pgfpathlineto{\pgfqpoint{4.053437in}{1.630366in}}%
\pgfpathlineto{\pgfqpoint{4.655708in}{1.690143in}}%
\pgfusepath{stroke}%
\end{pgfscope}%
\begin{pgfscope}%
\pgfpathrectangle{\pgfqpoint{0.439815in}{0.350111in}}{\pgfqpoint{4.215892in}{2.001772in}} %
\pgfusepath{clip}%
\pgfsetrectcap%
\pgfsetroundjoin%
\pgfsetlinewidth{2.007500pt}%
\definecolor{currentstroke}{rgb}{0.580392,0.403922,0.741176}%
\pgfsetstrokecolor{currentstroke}%
\pgfsetdash{}{0pt}%
\pgfpathmoveto{\pgfqpoint{0.429815in}{1.703406in}}%
\pgfpathlineto{\pgfqpoint{0.439815in}{1.703694in}}%
\pgfpathlineto{\pgfqpoint{1.042086in}{1.736527in}}%
\pgfpathlineto{\pgfqpoint{1.644356in}{1.795358in}}%
\pgfpathlineto{\pgfqpoint{2.246626in}{1.818644in}}%
\pgfpathlineto{\pgfqpoint{2.848897in}{1.855713in}}%
\pgfpathlineto{\pgfqpoint{3.451167in}{1.886670in}}%
\pgfpathlineto{\pgfqpoint{4.053437in}{1.931102in}}%
\pgfpathlineto{\pgfqpoint{4.655708in}{1.985474in}}%
\pgfusepath{stroke}%
\end{pgfscope}%
\begin{pgfscope}%
\pgfsetrectcap%
\pgfsetmiterjoin%
\pgfsetlinewidth{0.803000pt}%
\definecolor{currentstroke}{rgb}{0.000000,0.000000,0.000000}%
\pgfsetstrokecolor{currentstroke}%
\pgfsetdash{}{0pt}%
\pgfpathmoveto{\pgfqpoint{0.439815in}{0.350111in}}%
\pgfpathlineto{\pgfqpoint{0.439815in}{2.351883in}}%
\pgfusepath{stroke}%
\end{pgfscope}%
\begin{pgfscope}%
\pgfsetrectcap%
\pgfsetmiterjoin%
\pgfsetlinewidth{0.803000pt}%
\definecolor{currentstroke}{rgb}{0.000000,0.000000,0.000000}%
\pgfsetstrokecolor{currentstroke}%
\pgfsetdash{}{0pt}%
\pgfpathmoveto{\pgfqpoint{4.655708in}{0.350111in}}%
\pgfpathlineto{\pgfqpoint{4.655708in}{2.351883in}}%
\pgfusepath{stroke}%
\end{pgfscope}%
\begin{pgfscope}%
\pgfsetrectcap%
\pgfsetmiterjoin%
\pgfsetlinewidth{0.803000pt}%
\definecolor{currentstroke}{rgb}{0.000000,0.000000,0.000000}%
\pgfsetstrokecolor{currentstroke}%
\pgfsetdash{}{0pt}%
\pgfpathmoveto{\pgfqpoint{0.439815in}{0.350111in}}%
\pgfpathlineto{\pgfqpoint{4.655708in}{0.350111in}}%
\pgfusepath{stroke}%
\end{pgfscope}%
\begin{pgfscope}%
\pgfsetrectcap%
\pgfsetmiterjoin%
\pgfsetlinewidth{0.803000pt}%
\definecolor{currentstroke}{rgb}{0.000000,0.000000,0.000000}%
\pgfsetstrokecolor{currentstroke}%
\pgfsetdash{}{0pt}%
\pgfpathmoveto{\pgfqpoint{0.439815in}{2.351883in}}%
\pgfpathlineto{\pgfqpoint{4.655708in}{2.351883in}}%
\pgfusepath{stroke}%
\end{pgfscope}%
\begin{pgfscope}%
\pgfsetbuttcap%
\pgfsetmiterjoin%
\definecolor{currentfill}{rgb}{1.000000,1.000000,1.000000}%
\pgfsetfillcolor{currentfill}%
\pgfsetfillopacity{0.800000}%
\pgfsetlinewidth{1.003750pt}%
\definecolor{currentstroke}{rgb}{0.800000,0.800000,0.800000}%
\pgfsetstrokecolor{currentstroke}%
\pgfsetstrokeopacity{0.800000}%
\pgfsetdash{}{0pt}%
\pgfpathmoveto{\pgfqpoint{2.901374in}{0.405666in}}%
\pgfpathlineto{\pgfqpoint{4.577930in}{0.405666in}}%
\pgfpathquadraticcurveto{\pgfqpoint{4.600152in}{0.405666in}}{\pgfqpoint{4.600152in}{0.427889in}}%
\pgfpathlineto{\pgfqpoint{4.600152in}{1.198221in}}%
\pgfpathquadraticcurveto{\pgfqpoint{4.600152in}{1.220443in}}{\pgfqpoint{4.577930in}{1.220443in}}%
\pgfpathlineto{\pgfqpoint{2.901374in}{1.220443in}}%
\pgfpathquadraticcurveto{\pgfqpoint{2.879152in}{1.220443in}}{\pgfqpoint{2.879152in}{1.198221in}}%
\pgfpathlineto{\pgfqpoint{2.879152in}{0.427889in}}%
\pgfpathquadraticcurveto{\pgfqpoint{2.879152in}{0.405666in}}{\pgfqpoint{2.901374in}{0.405666in}}%
\pgfpathclose%
\pgfusepath{stroke,fill}%
\end{pgfscope}%
\begin{pgfscope}%
\pgfsetrectcap%
\pgfsetroundjoin%
\pgfsetlinewidth{2.007500pt}%
\definecolor{currentstroke}{rgb}{0.121569,0.466667,0.705882}%
\pgfsetstrokecolor{currentstroke}%
\pgfsetdash{}{0pt}%
\pgfpathmoveto{\pgfqpoint{2.923597in}{1.137109in}}%
\pgfpathlineto{\pgfqpoint{3.145819in}{1.137109in}}%
\pgfusepath{stroke}%
\end{pgfscope}%
\begin{pgfscope}%
\pgftext[x=3.234708in,y=1.098221in,left,base]{\rmfamily\fontsize{8.000000}{9.600000}\selectfont Deep Learning}%
\end{pgfscope}%
\begin{pgfscope}%
\pgfsetrectcap%
\pgfsetroundjoin%
\pgfsetlinewidth{2.007500pt}%
\definecolor{currentstroke}{rgb}{1.000000,0.498039,0.054902}%
\pgfsetstrokecolor{currentstroke}%
\pgfsetdash{}{0pt}%
\pgfpathmoveto{\pgfqpoint{2.923597in}{0.980554in}}%
\pgfpathlineto{\pgfqpoint{3.145819in}{0.980554in}}%
\pgfusepath{stroke}%
\end{pgfscope}%
\begin{pgfscope}%
\pgftext[x=3.234708in,y=0.941665in,left,base]{\rmfamily\fontsize{8.000000}{9.600000}\selectfont Reinforcement Learning}%
\end{pgfscope}%
\begin{pgfscope}%
\pgfsetrectcap%
\pgfsetroundjoin%
\pgfsetlinewidth{2.007500pt}%
\definecolor{currentstroke}{rgb}{0.172549,0.627451,0.172549}%
\pgfsetstrokecolor{currentstroke}%
\pgfsetdash{}{0pt}%
\pgfpathmoveto{\pgfqpoint{2.923597in}{0.824443in}}%
\pgfpathlineto{\pgfqpoint{3.145819in}{0.824443in}}%
\pgfusepath{stroke}%
\end{pgfscope}%
\begin{pgfscope}%
\pgftext[x=3.234708in,y=0.785554in,left,base]{\rmfamily\fontsize{8.000000}{9.600000}\selectfont Representation Learning}%
\end{pgfscope}%
\begin{pgfscope}%
\pgfsetrectcap%
\pgfsetroundjoin%
\pgfsetlinewidth{2.007500pt}%
\definecolor{currentstroke}{rgb}{0.839216,0.152941,0.156863}%
\pgfsetstrokecolor{currentstroke}%
\pgfsetdash{}{0pt}%
\pgfpathmoveto{\pgfqpoint{2.923597in}{0.667888in}}%
\pgfpathlineto{\pgfqpoint{3.145819in}{0.667888in}}%
\pgfusepath{stroke}%
\end{pgfscope}%
\begin{pgfscope}%
\pgftext[x=3.234708in,y=0.628999in,left,base]{\rmfamily\fontsize{8.000000}{9.600000}\selectfont Semi-supervised Learning}%
\end{pgfscope}%
\begin{pgfscope}%
\pgfsetrectcap%
\pgfsetroundjoin%
\pgfsetlinewidth{2.007500pt}%
\definecolor{currentstroke}{rgb}{0.580392,0.403922,0.741176}%
\pgfsetstrokecolor{currentstroke}%
\pgfsetdash{}{0pt}%
\pgfpathmoveto{\pgfqpoint{2.923597in}{0.511777in}}%
\pgfpathlineto{\pgfqpoint{3.145819in}{0.511777in}}%
\pgfusepath{stroke}%
\end{pgfscope}%
\begin{pgfscope}%
\pgftext[x=3.234708in,y=0.472888in,left,base]{\rmfamily\fontsize{8.000000}{9.600000}\selectfont Unsupervised Learning}%
\end{pgfscope}%
\end{pgfpicture}%
}
\makeatother%
\endgroup%

    \caption{Publications by year referencing ML sub-fields according to Google Scholar search results.}
    \label{fig:pubs}
\end{figure}

Machine learning (ML) contains a number of increasingly popular sub-fields: reinforcement learning, representation learning, etc. The number of publications in each of these sub-fields alone, per year, can quickly become overwhelming to keep up with, constituting a flood of information as seen in Figure~\ref{fig:pubs}. In fact, according to a Google Scholar Search, the term "machine learning" showed up in approximately 106,000 works published online in 2017\footnote{\url{https://scholar.google.com/scholar?q=\%22machine+learning\%22&hl=en&as_sdt=0\%2C5&as_ylo=2017&as_yhi=2017}}. 
With this magnitude of yearly publications, several problems arise. First, it becomes nearly impossible for a single researcher to keep up with all the incoming papers on even one popular sub-field of machine learning. 
This has the potential to result in duplicated innovations and relevant papers being missed (and therefore left uncited and not compared against). 

With the large number of related publications, a need arises in distilling this flood of information into useful methodologies and lessons. 
This is particularly important due to the increasing use of machine learning in production systems, where we often seek to translate the most proven machine learning research to potentially high-stakes applications.
However, ML is not the first field to output an enormous and rapidly growing set of research results; medicine and epidemiology, for example, also have such large quantities of yearly publications.  
These other fields have taken to meta-analyses and systematic reviews on narrow subject matters to aggregate information. We believe it is well worth considering how and whether machine learning might similarly make use of meta-analyses and systematic reviews to support the best aggregation and building of scientific knowledge. 

\section{Background}

\subsection{Meta-Analysis}
In the early days of statistical inference and probability theory, Ronald Fisher -- a large contributor to what has become statistical significance testing -- suggested that ``experimental demonstration of, say, an interesting psychological phenomenon required confirmation by similar experiments in other laboratories''~\cite{gigerenzer1990empire}. That is, the use of significance testing alone in a single experimental setting was not enough to sufficiently accept a hypothesis. Similarly, \citet{simpson1904report} began combining different observation analyses to increase sample sizes that were deemed were not large enough~\cite{simpson1904report,o2007historical}. That work can be considered one of the earliest known uses of meta-analysis, which we will define here according to~\citet{haidich2010meta} as: ``a quantitative, formal, epidemiological study design used to systematically assess previous research studies to derive conclusions about that body of research.'' These themes of repeated experiments in different settings as a foundation of inference have become a backbone in medicine and epidemiology among other fields, such that \citet{haidich2010meta} place it as the highest tier of evidence and \citet{gurevitch2018meta} describe these methods as essential to scientific research.
In ML research, meta-analyses have not yet been popularized (possibly due to several reasons to be discussed in Section \ref{sec:problems}). However, there have already been several works which either perform meta-analyses or something similar -- though perhaps not under the same formality as clinical research~\cite{chattopadhyay2018comprehensive,gomez2017meta,melis2017state,henderson2017deep}.

\subsection{Systematic Review}
\citet{haidich2010meta} provide an excellent definition of systematic reviews as well. They state:
\begin{quote}
    Meta-analyses are a subset of systematic review. A systematic review attempts to collate empirical evidence that fits prespecified eligibility criteria to answer a specific research question. The key characteristics of a systematic review are a clearly stated set of objectives with predefined eligibility criteria for studies; an explicit, reproducible methodology; a systematic search that attempts to identify all studies that meet the eligibility criteria; an assessment of the validity of the findings of the included studies (e.g., through the assessment of risk of bias); and a systematic presentation and synthesis of the attributes and findings from the studies used.
\end{quote}

 While, again, ML research publishes a number of reviews and survey papers (e.g., \citet{lisboa2006use,li2017deep,kober2013reinforcement}), often, these reviews are not conducted in a systematic way as described by \citet{haidich2010meta}.

\section{Meta-Analysis and Systematic Review in ML Literature}
Other fields have fairly straightforward usages of meta-analysis and systematic review and potential guidelines, such as those presented by \citet{haidich2010meta}, \citet{gopalakrishnan2013systematic}, and \citet{muller2017ten}. However, using such methods in ML may not be straightforward. For example, what sorts of questions can we answer via meta-analyses and systematic reviews? What would be a parallel paradigm to meta-analyses in medicine? Additionally, are enough tools and information provided in the literature to perform these analyses for ML research? While these are still open questions -- as the meta-analysis and systematic review literature in ML research is young (if it can be classified as existing) -- we will suggest potential uses, benefits, and problems with using such methods in ML research here. We also note that our discussion pertains primarily to algorithmic and empirical ML research, rather than theoretical ML, where researchers are showing the utility and benefits of an algorithm on a set of datasets or domains.

\subsection{Potential Uses and Benefits}

First, we will look at what kinds of questions can these methods be used to investigate. 
In ML discussions, a number of questions have arisen in recent years which -- for better or for worse -- have resulted in serious debates. These types of questions include debates about whether connectionist (deep) learning is limited in its ability to advance machine learning, among others (see \citet{chauvet201830} for examples of cyclical debates in ML research).
These types of questions likely contain varying levels of subjectivity and even meta-analyses may not be able to resolve them. 

There are, however, some particular properties of ML research that make the field well suited for more narrow questions that 
may be addressed by a meta-analysis or systematic review. 

For example, most ML research involves comparing a new algorithm against one or more baselines. 
Often, similar baselines are re-used across a variety of works on different datasets and domains (possibly with different implementations). Using systematic review and meta-analysis, we can begin to answer questions about baseline algorithms, such as how stable is performance across implementations (a proxy for how easy to implement the core algorithm is) and how does performance vary across properties of datasets/domains and hyperparameters? 
In fact, meta-analysis of the baselines would likely provide a somewhat lower bound on performance due to incentive structures in modern research~\cite{sculley2018winner}. That is, there is incentive to beat the baseline and find conditions where baselines perform poorly. 

Using these methods, we can also begin to systematically answer questions about particular techniques and comparisons between techniques.  
For example, meta-analysis could help answer questions such as: does batch normalization tend to improve results in random samples of experiments in various contexts and what is the magnitude of its effect?
Of course, the current field tends to naturally perform these analyses through ablation studies,
but a meta-analysis may reveal gaps in the analyses that the current literature provides and moreover can show trends in aggregate across implementations and scenarios.
This is particularly important as other works have shown how different codebases can yield different results under similar conditions~\cite{henderson2017deep}. 
Meta-analyses and systematic reviews can show trends that no individual work can show.

\subsubsection{Inference and Synthesis of Aggregate Knowledge}

\citet{rahimi2017} compared current ML research to alchemy in their acceptance of the Test of Time Award at the Neural Information Processing Systems Conference (NeurIPS) in 2017. 
Alchemy in time turned into chemistry through systematic inference and deduction of effects, and in most other scientific areas systematic review and meta-analysis have played a large role in this over the last half-century~\cite{o2007historical}.
Moreover, such methods allow us to aggregate knowledge such that even the most novice researchers can quickly bootstrap their research by seeing quantitatively and systematically what are the effects of using different techniques. 

For example, it may help us determine in aggregate which settings can benefit most from temporal abstraction or what classes of applications may benefit from model-free algorithms versus model-based algorithms. Furthermore, it may avoid scenarios where discoveries are lost and repeated many times. This can guide future research toward new developments in a more efficient manner. Conversely, if systematic review and meta-analysis yields that an algorithm baseline severely under-preforms, it can safely be replaced to save on computational complexity of future experimentation on that baseline. 

\noindent \textbf{Finding Gaps in Knowledge Quickly} Systematic research can also help us find gaps in knowledge and evaluation methodologies quickly. If a systematic review and meta-analysis cannot make concrete inferences on a subject matter, it is likely that not enough information was provided (either through a lack of ablation experimentation or through inadequate evaluation methods). Thus, these gaps can be quickly closed without waiting until the community organically comes to such realizations.

\noindent \textbf{Prescriptiveness} The ability to systematically compare phenomena and algorithms in particular conditions is incredibly valuable for application development. Such analyses can provide prescriptions to engineers and applied researchers attempting to use state-of-the-art methods without wasting time on methods which haven't shown consistency in a meta-analysis. This directly parallels the adoption of methodologies in medical research.

\subsection{Problems}
\label{sec:problems}
As with any statistical inference technique, if not used properly, meta-analysis can contain methodological issues of its own -- and as a result have yielded some criticism~\cite{oaks1986statistical,doi:10.1093/humrep/deu127,ioannidis2016mass,eysenck1994systematic}. These issues can include sampling biases, improper inference of statistical significance, improper combination of results, not well-defined criteria for inclusion, etc. However, it can be argued that this is a matter of execution rather than idea.

Another argument may be that the field already naturally generates pseudo-meta-analyses and does the job of systematic review, such as~\cite{melis2017state,henderson2017deep} among others. However, these studies may not bear the marks of systematic selection and application as prescribed by meta-analysis methodology in other fields which help account for sampling biases and previously mentioned confounding factors. Such studies are at their core ablation analyses, not systematic meta-analyses and, while equally valuable, address slightly different problems.

However, one might argue that such ablation studies are adequate without need for meta-analysis and systematic review. To this, we can refer back to the notion of reproducibility as defined by~\citet{result2017artifact}. If a method can be reproduced successfully across codebases between different experimental teams, this adds a level of trust in the method that can be extracted by meta-analyses. That is, the original definition of the work was adequate enough to yield similar results across a variety of implementation conditions (something useful for transfer to real-world development of applications using the novel methods). 

\subsubsection{Lack of Publication Venues}
One major issue with encouraging meta-analyses is the lack of venues and incentives to publish such studies. These studies may be rejected in current conferences and journals for a lack of novelty, especially if they confirm previously known phenomena, though they may be useful for inference and synthesis of aggregate knowledge. 
We note that there are some exceptions, such as the \textit{Foundations and Trends in Machine Learning} journal, but these journals generally accept surveys or tutorials rather than systematic reviews or meta-analyses.
Providing reliable conference and journal venues similarly to other fields may yield more incentive structures for publication of systematic reviews and meta-analyses in ML research. 
In other fields, dedicated journals exist specifically for systematic reviews and meta-analyses (e.g., the \textit{Nature Reviews} series of journals). While reviews can appear in some machine learning journals, they remain rare compared to other fields. 
Perhaps a bigger barrier is that machine learning, like many other engineering disciplines, primarily values new innovations over careful synthesis of prior results. This is changing with more analysis-like works being released (e.g.,~\citet{henderson2017deep} and \citet{shallue2018measuring}). 
Yet, for meta-analyses and systematic reviews to take hold in machine learning, such contributions will need to receive support as an important type of publication that can help move the field further.

\section{Conclusion}
We have proposed promoting the paradigm of systematic review and meta-analysis in ML research. Such a paradigm has not yet taken hold in our field, yet may be a necessity for distilling knowledge from the current flood of novel work. While there may be positives and negatives to using these methods, we emphasize that experimentation with these methods of research may be of benefit to the community. Through this work, we would like to draw attention to the existence of these methods, encourage debate on them, and in general encourage their use in ML research. Other fields have already developed these methods extensively~\cite{o2007historical,gurevitch2018meta,gopalakrishnan2013systematic,muller2017ten}, so it is up to us in the ML community to draw from their experience and determine -- in aggregate -- what we do and don't know about ML. 

\bibliographystyle{plainnat}

\end{document}